\documentclass[preprint,showpacs,preprintnumbers,amsmath,amssymb]{revtex4}

\usepackage{graphicx}% Include figure files
\usepackage{dcolumn}% Align table columns on decimal point
\usepackage{bm}% bold math

\begin{document}

\title{Optimal Location of Two Laser-interferometric Detectors for Gravitational Wave 
Backgrounds at 100 $\rm{MHz}$}

\author{Atsushi Nishizawa$^{1*}$, Seiji Kawamura$^2$, Tomotada Akutsu$^3$,\\ Koji Arai
$^2$, Kazuhiro Yamamoto$^2$, Daisuke Tatsumi$^2$, Erina Nishida$^4$, \\
Masa-aki Sakagami$^1$, Takeshi Chiba$^5$, Ryuichi Takahashi$^6$, Naoshi Sugiyama$^6$}

\affiliation{$^1$ Graduate School of Human and Environmental Studies, Kyoto University, 
Kyoto 606-8501, Japan}
\email{atsushi.nishizawa@nao.ac.jp}
\affiliation{$^2$ TAMA Project, National Astronomical Observatory of Japan, Mitaka, 
Tokyo 181-8588, Japan}
\affiliation{$^3$ Department of Astronomy, School of Science, University of Tokyo, 
Bunkyo-ku, Tokyo 113-0033, Japan}
\affiliation{$^4$ Ochanomizu University, Bunkyo-ku, Tokyo 112-8610, Japan }
\affiliation{$^5$ Department of Physics, College of Humanities and Sciences, Nihon 
University, Tokyo 156-8550, Japan}
\affiliation{$^6$ Graduate School of Science, Nagoya University, Nagoya 467-8602, 
Japan}

\date{\today}

\begin{abstract}
Recently, observational searches for gravitational wave background (GWB) have been
developed and given constraints on the energy density of GWB in a 
broad range of frequencies. These constraints have already resulted in the rejection of some theoretical models of relatively large GWB spectra. However, at $100 \, \rm{MHz}$, there is no strict upper 
limit from {\it{direct}} observation, though an {\it{indirect}} limit exists due to $^4\rm{He}$ 
abundance due to big-bang nucleosynthesis.
In our previous paper, we investigated the detector designs that can effectively respond to GW at high frequencies, where the wavelength of GW is comparable to the size of a detector, and found that the configuration, a so-called synchronous-recycling interferometer is best at these sensitivity. 
In this paper, we investigated the optimal location of two synchronous-recycling interferometers and derived their cross-correlation sensitivity to GWB. We found that the sensitivity is nearly optimized and hardly changed if two coaligned detectors are located in a range $\pm 0.2\, \rm{m}$, and that the sensitivity achievable in an experiment is $h_{100}^2 \Omega_{\rm{gw}} \approx 1.4 \times 10^{14}$. This would be far below compared with the constraint previously obtained in experiments.
\end{abstract}

%\pacs{Valid PACS appear here}
\maketitle

\section{Introduction}
There are many theoretical predictions of gravitational wave background (GWB) in a 
broad range of frequencies, $10^{-18}-10^{10}\, \rm{Hz}$. Some models in cosmology 
and particle physics predict relatively large stochastic
GWB at ultra high frequency $\sim 100 \, \rm{MHz}$; the quintessential inflation model proposed by \cite{bib1}, in which a large GWB spectrum is produced during the kinetic energy-dominated era after the inflationary expansion of the universe \cite{bib2,bib3,bib4,bib5}, the violent reheating process after inflation, a so-called preheating \cite{bib6,bib7,bib8}, pre-big-bang scenarios in string cosmology \cite{bib9,bib10,bib11}, the binary 
evolution and coalescence of primordial black holes produced in 
the early universe \cite{bib12,bib13,bib29} and their evaporation \cite{bib14}. Recent 
predictions of GW emission from black strings in the Randall-Sundrum model also generate 
spectral features characteristic of the curvature of extra dimensions at high frequencies 
\cite{bib15,bib16}. For the inquiry into high energy physics, testing these models with gravitational wave (GW) detectors for high frequencies is very important. 

Upper limits on GWB in wide-frequency ranges have been obtained from various 
observations; cosmic microwave radiation at $10^{-18}-10^{-15} \,\rm{Hz}$ \cite{bib17}, 
pulsar timing at $10^{-9}-10^{-7}\, \rm{Hz}$ \cite{bib18}, Doppler tracking of the {\it
{Cassini}} spacecraft at $10^{-6}-10^{-3}\, \rm{Hz}$ \cite{bib19}, direct observation by 
LIGO at $10-10^{4} \rm{Hz}$ \cite{bib20}, $^4\rm{He}$ abundance due to big-bang 
nucleosynthesis at frequencies greater than $10^{-10}\, \rm{Hz}$ \cite{bib21}. 
Nevertheless, as far as we know, no direct experiment has been done above $10^5 \, 
\rm{Hz}$ except for the experiment by A. M. Cruise and R. M. J. Ingley \cite{bib22}. They 
have used electromagnetic waveguides and obtained an upper limit on the amplitude of 
GW backgrounds, $h\leq 10^{-14}$ corresponding to $ h_{100}^2\Omega_{\rm{gw}} \leq 
10^{34}$ at $100 \, \rm{MHz}$, where $h_{100}$ is the Hubble constant normalized with 
$100 \,\rm{km}\,\rm{sec}^{-1}\,\rm{Mpc}^{-1}$ and $\Omega_{\rm{gw}}$ is the energy 
density of GWB per logarithmic frequency bin normalized by the critical energy density of 
the universe, that is,
\begin{equation}
\Omega_{\rm{gw}}(f) = \frac{1}{\rho_c}\frac{d\rho_{\rm{gw}}}{d \ln f} \:.
\label{eq5}
\end{equation}
This constraint is much weaker than the constraints at other frequencies. Therefore, a 
much tighter boundary above $10^5 \, \rm{Hz}$ is needed to test various theoretical models.

For our purpose of detecting GWB at $100\,\rm{MHz}$, a detector with narrow 
bandwidth and high sensitivity is sufficient because we do not intend to know the exact waveform. In our previous paper \cite{bib23}, we considered three detector 
designs: a synchronous-recycling 
interferometer, a Fabry-Perot Michelson interferometer and an L-shaped cavity Michelson 
interferometer, and investigated their GW responses. As a result, we found that the 
synchronous recycling interferometer (SRI) \cite{bib24} is the most sensitive at $100\rm
{MHz}$, in which the angular mean sensitivity is slightly worse than that in low frequencies because the detector size is comparable to GW wavelength.

To detect GWB with smaller amplitude than detector noise, one has to correlate signals from two detectors in order to distinguish the GW signal. The analytical method has been well developed by several authors \cite
{bib25,bib26,bib27}. In these references, they assume that GW wavelength is much 
larger than the detector size, which is a so-called long-wave approximation. However, it is 
not valid in our situation around $100\,\rm{MHz}$, in which the GW wavelength is 
comparable to the detector size. This means the relative location of the two SRIs significantly affects the correlation sensitivity to GWB and the response
function of a detector should be taken into account properly. Thus, in this paper, we will extend the previous analytical method of correlation, including the response functions of the detectors, and investigate the dependence of the sensitivity on the relative location of two detectors.

This paper is organized as follows. In Sec. II, we will briefly review the correlation analysis for GWB and extend it including a full detector-response function. In Sec. III, we will explain a synchronous recycling interferometer and its response function. Sec. IV is the main section of this paper and is devoted to the evaluation of an overlap reduction function and the consideration of dependence of sensitivity on the locations of detectors. We will calculate the sensitivity of two SRIs to GWB in Sec. V and conclude in Sec. VI.

\section{Correlation analysis}
\label{sec2}
In this section, we will briefly review the formalism of correlation analysis for GWB \cite
{bib21,bib25} and redefine a part of the formalism, including detector response functions, which are necessary to take the finite size of detectors into account.

Let us consider the outputs of a detector, $s(t)=h(t)+n(t)$, where $h(t)$ and $n(t)$ are the GW signal and the noise of a detector. At generic point $\vec{\mathbf{X}}$, the gravitational metric 
perturbations in the transverse traceless gauge are given by
\begin{equation}
\mathbf{h}(t,\vec{\mathbf{X}})=\sum_p \int _{S^2} d\hat{\mathbf{\Omega}} \int_{-\infty}^
{\infty}df\, \tilde{h}_p (f, \hat{\mathbf{\Omega}})\, e^{2\pi if (t-\hat{\mathbf{\Omega}} \cdot 
\vec{\mathbf{X}}/c)}\, \mathbf{e}_p(\hat{\mathbf{\Omega}})\:,
\label{eq2}
\end{equation}
where $c$ is speed of light, $\hat{\mathbf{\Omega}}$ is a unit vector directed at GW 
propagation and $\tilde{h}_p(f, \hat{\mathbf{\Omega}})$ is the Fourier transform of GW amplitude with polarizations $p=+, \times$. Polarizarion tensors $\mathbf{e}_p(\hat{\mathbf{\Omega}})$, can be written as 
\begin{eqnarray}
\mathbf{e}_{+}(\hat{\mathbf{\Omega}}) &\equiv & \hat{\mathbf{m}} \otimes \hat{\mathbf
{m}} -\hat{\mathbf{n}} \otimes \hat{\mathbf{n}}, \nonumber \\
\mathbf{e}_{\times}(\hat{\mathbf{\Omega}}) &\equiv & \hat{\mathbf{m}} \otimes \hat
{\mathbf{n}} +\hat{\mathbf{n}} \otimes \hat{\mathbf{m}}\:, \nonumber 
\end{eqnarray}
where the unit vectors $\hat{\mathbf{m}}$, $\hat{\mathbf{n}}$ are orthogonal to $\hat
{\mathbf{\Omega}}$ and to each other. 

In this paper, we assume that GWB is  (i) isotropic, (ii) unpolarized, (iii) stationary, and (iv) Gaussian, and (v) has small amplitude compared with that of noise, $|h(t)|\ll |n(t)|$. These assumptions (i) - 
(iv) are discussed in \cite{bib25}, and (v) is expected at high frequencies around 
$100\,\rm{MHz}$ because there exists an indirect upper limit on GWB by big-bang 
nucleosynthesis. These assumptions (i) - (iv) are expressed by
\begin{equation}
\langle \tilde{h}_p^{\ast} (f,\hat{\Omega} ) \tilde{h}_{p^{\prime}} (f^{\prime},\hat{\Omega^
{\prime}} ) \rangle \equiv \delta(f-f^{\prime})\frac{1}{4\pi} \delta^2 ( \hat{\Omega},\hat
{\Omega^{\prime}}) \delta_{pp^\prime} \frac{1}{2} S_h (f),
\label{eq6}
\end{equation}
where $\delta^2 ( \hat{\Omega},\hat{\Omega^{\prime}}) \equiv \delta(\phi-\phi^{\prime}) 
\delta (\cos\theta - \cos\theta^\prime )$, $S_h(f)$ is the one-sided power spectral density and is defined by Eq. (\ref{eq6}), and $\langle \cdots \rangle$ denotes ensemble average. The power spectral density $S_h(f)$ is related to $\Omega_{\rm{gw}}$ by
\begin{equation}
\Omega_{\rm{gw}} (f) = \left( \frac{4\pi^2}{3H_0^2} \right) f^3 S_h (f)\:,
\label{eq7}
\end{equation}                                                          
where $H_0$ is the Hubble constant \cite{bib21}. 

GW signal $h(t)$ from a detector is given by $\mathbf{D}(f, \hat{\mathbf{\Omega}}):
\mathbf{h}(f, \hat{\mathbf{\Omega}})$, where the symbol : denotes contraction between 
tensors, and $\mathbf{D}(f, \hat{\mathbf{\Omega}})$ is a so-called detector tensor, which 
describes the total response of a detector and maps the gravitational metric perturbation
to the GW signal from a detector. We define it including detector response functions as 
\begin{equation}
\mathbf{D}(f, \hat{\mathbf{\Omega}}) \equiv \frac{1}{2}\left[ (\hat{\mathbf{u}} \otimes \hat
{\mathbf{u}}) {\cal{T}}(f, \hat{\mathbf{\Omega}} \cdot \hat{\mathbf{u}}) -(\hat{\mathbf{v}}
\otimes \hat{\mathbf{v}}) {\cal{T}}(f, \hat{\mathbf{\Omega}} \cdot \hat{\mathbf{v}})\right]\:.
\label{eq1}
\end{equation}
Here $\hat{\mathbf{u}}$ and $\hat{\mathbf{v}}$ are unit vectors. We assume that they are orthogonal to each other and are directed to each detector arm.
The function ${\cal{T}}$ is a detector response function, which we will introduce below,
that describes the effect of finite arm length on propagating light and gives unity in low 
frequency limit. In the detector whose arm length is much smaller than the 
wavelength of GW, this function is approximated to unity, while in our detector whose 
detector size is comparable to GW wavelength, the function significantly affects the 
response of the detector. Using Eqs. (\ref{eq2}) and (\ref{eq1}), GW signal $h(t)$ can be 
written as
\begin{equation}
h(t,\vec{\mathbf{X}})=\sum_p \int _{S^2} d\hat{\mathbf{\Omega}} \int_{-\infty}^{\infty}df\, 
\tilde{h}_p (f, \hat{\mathbf{\Omega}})\, e^{2\pi if (t-\hat{\mathbf{\Omega}} \cdot \vec
{\mathbf{X}}/c)} F_p(f, \hat{\mathbf{\Omega}})\:,
\label{eq3}
\end{equation}
where an angular pattern function of a detector $F_p(f, \hat{\mathbf{\Omega}})$ is 
defined by 
\begin{equation}
F_p(f, \hat{\mathbf{\Omega}}) \equiv \mathbf{D}(f, \hat{\mathbf{\Omega}}) : \mathbf{e}_p
(\hat{\mathbf{\Omega}})\:.
\end{equation}

Cross-correlation signal $S$ between two detectors is defined as
\begin{equation}
S \equiv \int_{-T/2}^{T/2} dt \int_{-T/2}^{T/2} dt^{\prime}\, s_1(t) s_2(t^{\prime} ) Q(t-t^
{\prime}),
\end{equation}
where $s_1$ and $s_2$ are an output from each detector, $T$ is observation time. $Q(t-t^{\prime})$ is an arbitrary real function, which is called an optimal filter. We will 
determine its form below so that signal-to-noise ratio (SNR) is maximized. Fourier transforming $s_1(t)$ and $s_2(t)$, one can obtain
\begin{equation}
S=\int_{-\infty}^{\infty}df \int_{-\infty}^{\infty}df^{\prime} \delta _T (f-f ^{\prime}) \tilde{s}_1^
{*}(f) \tilde{s}_2(f^{\prime}) \tilde{Q}(f^{\prime}),
\label{eq9}
\end{equation}
where $\tilde{s}_1(f)$, $\tilde{s}_2(f)$ and $\tilde{Q}(f)$ are the Fourier transforms of 
$s_1(t)$, $s_2(t)$ and $Q(t-t^{\prime})$, respectively. $\delta_T(f)$ is the finite-time approximation to the Dirac 
delta function defined by
\begin{equation}
\delta _T (f)\equiv \int_{-T/2}^{T/2} dt \,e^{-2\pi ift}=\frac{\sin(\pi fT)}{\pi f}\:.
\nonumber
\end{equation}
In the above derivation, we took the limit of large $T$ for one of the integrals. This is justified by the fact that, in general, $Q(t-t^{\prime})$ rapidly decreases for large $|t-t^{\prime}|$.
The correlation signal obtained above ideally has a contribution from only the GW signal since 
we assume that noise has no correlation between two detectors. Thus, we take 
ensemble average of Eq. (\ref{eq9}) and obtain the signal from GWB,
\begin{equation}
\mu \equiv \langle S \rangle =\int_{-\infty}^{\infty}df \int_{-\infty}^{\infty}df^{\prime} \delta_T(f-f^{\prime}) \langle \tilde{h}_1^{*}(f) \tilde{h}_2(f^{\prime}) 
\rangle \tilde{Q}(f^{\prime})\:.
\label{eq4}
\end{equation}
Substituting the Fourier transform of Eq. (\ref{eq3}),
\begin{equation}
\tilde{h}(f)=\sum_p \int_{S^2}d\hat{\mathbf{\Omega}}\, \tilde{h}_p(f,\hat{\mathbf
{\Omega}} )e^{-2\pi if \hat{\mathbf{\Omega}} \cdot \vec{\mathbf{X}}/c} F^p (f, \hat{\mathbf
{\Omega}})\:,
\label{eq17}
\end{equation}
into Eq. (\ref{eq4}), and using Eqs. (\ref{eq6}) and (\ref{eq7}), one can obtain
\begin{equation}
\mu= \frac{3H_0^2}{20\pi^2}T\int_{-\infty}^{\infty} df |f|^{-3} \Omega_{\rm{gw}}(|f|)\, 
\gamma (|f|)\, \tilde{Q}(f).
\label{eq10}
\end{equation}
Here we defined the overlap reduction function,
\begin{equation}
\gamma (f) \equiv \frac{5}{8\pi} \sum_p \int_{S^2} d\hat{\mathbf{\Omega}} \;e^{2\pi if \hat
{\mathbf{\Omega}} \cdot \Delta \vec{\mathbf{X}}/c } F_1^{p\; \ast}(f, \hat{\mathbf
{\Omega}}) F_2^p(f, \hat{\mathbf{\Omega}})\:,
\label{eq8}
\end{equation}
where the separation of two detectors is $\Delta \vec{\mathbf{X}} \equiv \vec{\mathbf{X}}
_1-\vec{\mathbf{X}}_2$. The factor in front of Eq. (\ref{eq8}) is a normalization and the 
overlap reduction function gives unity in low frequency limit. This definition is slightly different 
from that in other papers \cite{bib21,bib25}. In the references, the overlap reduction 
function is defined as meaning how GW signals in two detectors are correlated, and 
equals unity for colocated and coaligned detectors, while our definition includes the 
detector response functions ${\cal{T}}$, as one can see explicitly in Eq. (\ref{eq1}), and does not 
give unity even for colocated and coaligned detector at high frequencies. Namely, we regard the loss of GW 
signals of detectors as the reduction of overlap between two detectors. We will see below that this is important for the detector whose size is comparable to GW wavelength.

Next, we will calculate the variance of a correlation signal. In this paper, we assume that noises in two detectors do not correlate at all and that the magnitude of GW 
signal is much smaller than that of noise. Consequently, the variance of correlation signal is
\begin{equation}
\sigma ^2 \equiv  \langle S^2 \rangle - \langle S\rangle ^2 \approx \langle S^2 \rangle\:.
\nonumber
\end{equation}  
Then, using Eq. (\ref{eq9}), it follows
\begin{eqnarray}
\sigma ^2&\approx& \int_{-\infty}^{\infty} df \int_{-\infty}^{\infty} df^{\prime}\, \tilde{Q}(f) \tilde{Q}^{\ast}(f^{\prime})\, \langle 
\tilde{s}^{\ast}_1(f) \tilde{s}_1(f^{\prime}) \rangle \, \langle \tilde{s}_2(f) \tilde{s}^{\ast}_2
(f^{\prime}) \rangle \nonumber \\
&\approx &\frac{T}{4} \int_{-\infty}^{\infty} df \,P_1(|f|) P_2(|f|)\, | \tilde{Q}(f) | ^2\:,
\label{eq11}
\end{eqnarray}
where the one-sided power spectrum density of noise is defined by
 \begin{equation}
\langle \tilde{n}^{\ast}_i(f) \tilde{n}_i(f^{\prime}) \rangle \equiv \frac{1}{2}\delta (f-f^
{\prime})P_i(f) \nonumber, \;\;\;\;\;\;\;\; i=1,2\:.
\end{equation}

Now let us determine the form of the optimal filter $\tilde{Q}(f)$. Equations (\ref{eq10}) 
and (\ref{eq11}) are expressed more simply, using an inner product
\begin{equation}
(A, B) \equiv \int_{-\infty}^{\infty} df A^{\ast}(f) B(f) P_1(|f|) P_2(|f|)\:,
\nonumber
\end{equation} 
as
\begin{eqnarray}
\mu &=& \frac{3H_0^2}{20\pi^2}T \left( \tilde{Q}, \frac{\gamma (|f|) \Omega_{\rm{gw}}(|f|)}{|
f|^3 P_1(|f|) P_2(|f|)} \right)\:,
\label{eq12} \\
\sigma ^2 &\approx & \frac{T}{4} \left( \tilde{Q}, \tilde{Q} \right)\:.
\label{eq13} 
\end{eqnarray}
From Eqs. (\ref{eq12}) and (\ref{eq13}), SNR for GWB is defined as
${\rm{SNR}} \equiv \mu / \sigma $. Therefore, the optimal filter function turns out to be
\begin{equation}
\tilde{Q}(f)=\lambda \,\frac{\gamma (f) \Omega_{\rm{gw}}(|f|)}{|f|^3 P_1(|f|) P_2(|f|)},
\nonumber
\end{equation} 
with an arbitrary normalization factor $\lambda$. Applying this optimal filter to the above 
equations, we obtain maximal SNR
\begin{eqnarray}
{\rm{SNR}} &=& \frac{3H_0^2}{10\pi^2 } \sqrt{T} \left[ \int_{-\infty}^{\infty} df \frac{\gamma ^2 (|f|) 
\Omega ^2_{\rm{gw}}(|f|)}{f^6 P_1(|f|) P_2(|f|)} \right]
^{1/2} \nonumber \\
&\approx &3.19 \times 10^{-37} \sqrt{T} \left[ \int_{-\infty}^{\infty} df \frac{ \gamma ^2 (|f|) 
\{h_{100}^2 \Omega _{\rm{gw}}(|f|) \} ^2 }{f^6 P_1(|f|) P_2(|f|)} \right]^{1/2} \:.
\label{eq18}
\end{eqnarray}
In the transformation from the first to the second, we used $H_0= 100\,h_{100}\, {\rm
{km}\,\rm{s}^{-1}\,\rm{Mpc}}^{-1} \approx 3.24\, h_{100} \times 10^{-18}\,{\rm{s}}^{-1}$.
 
\section{synchronous-recycling interferometer}
\begin{figure}[h]
\begin{center}
\includegraphics[width=8cm]{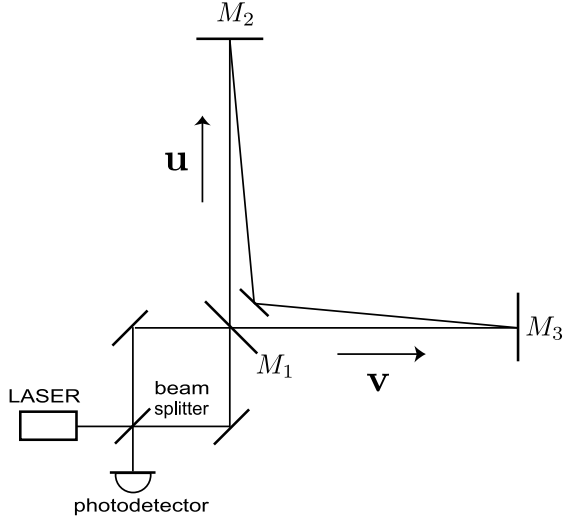}
\caption{Design of synchronous recycling interferometer (SRI).}
\label{fig1}
\end{center}
\end{figure}

In this section, we will describe the detector response function ${\cal{T}}(f,\hat{\mathbf
{\Omega}})$ and spectrum density of noise $P_i,\:i=1,2$ of a synchronous-recycling 
interferometer (SRI), which are needed to calculate the sensitivity to GWB. The design of SRI 
is shown in Fig. \ref{fig1}, which was first proposed by R. W. P. Drever in \cite{bib24}. 
Laser light is split at a beam splitter and is sent into a synchronous-recycling cavity through a recycling mirror, which is the mirror $M_1$ in Fig. \ref{fig1}. The beams circulating clockwise and 
counterclockwise in the cavity experience GWs and mirror displacements, leave the 
cavity, and are recombined at the beam splitter. Then, the differential signal is detected 
at a photodetector. The advantage of SRI is that GW signals at certain frequencies are 
accumulated and amplified because the light beams experience GWs with the same sign of phases during round trips in the folded cavity. 
Consider GW propagating normally to the detector plane with an optimal polarization. In 
this case, the GW signal is amplified at the frequencies $f = (2n-1) \times c/4L,\;n=1,2,
\cdots $, where $L$ is the arm length \footnote{We assume the
Sagnac part in front of the cavity is much smaller than the cavity.}. The disadvantage of SRI is less sensitivity for GWs at low frequencies, $f < c/4L$, 
because the GW signal is integrated in the cavity and canceled out as the frequency decreases.

\begin{figure}[h]
\begin{center}
\includegraphics[width=8cm]{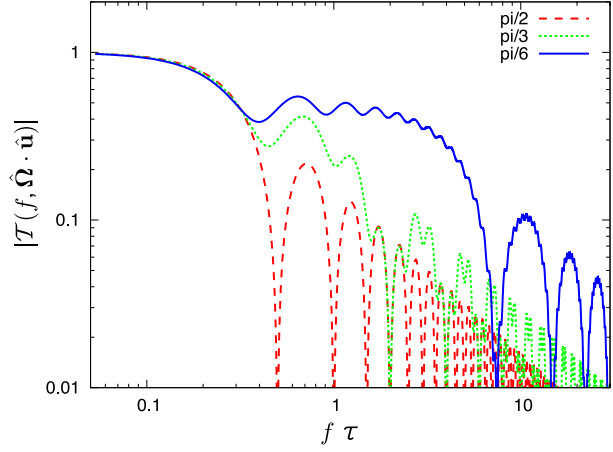}
\caption{(color online) Arm response functions as a function of $f \tau$. Each plot is for $\arccos(\hat{\mathbf{\Omega}}\cdot \hat{\mathbf{u}})= \pi /2,\,\pi/3,\,\pi/6$, respectively, as shown by plot labels in the figure.}
\label{fig6}
\end{center}
\end{figure}

In our previous paper \cite{bib23}, we derived the GW response of an SRI and found that 
it can be written in the form $\tilde{\delta \Phi }(f,\,\hat{\mathbf{\Omega}})=\alpha (f)\, 
\tilde{\delta \phi} (f,\,\hat{\mathbf{\Omega}})$, where $\tilde{\delta \phi} (f,\,\hat{\mathbf
{\Omega}})$ denotes the Fourier component of phase shift due to GW during the round 
trip of light in a recycling cavity and $\alpha (f)$ denotes an optical amplification factor of 
light in the cavity. The concrete expressions are written as
\begin{equation}
\alpha (f) =-\frac{R_E T_F^2}{(R_F-R_E)(1-R_F R_E \;e^{-8 \pi i f \tau})}\;, 
\end{equation}
\begin{eqnarray}
\tilde{\delta \phi}(f, \hat{\mathbf{\Omega}} ) &=& (1-e^{-4\pi i f \tau})  \frac{\omega}{2\pi 
f}\;e^{-2\pi i f (\tau+\hat{\mathbf{\Omega}} \cdot \vec{\mathbf{X}}/c)} \sum_p \tilde{h}_p(f, \hat{\mathbf{\Omega}} )\, \mathbf{e}_p (\hat{\mathbf{\Omega}} ) \nonumber \\
&&: \biggl[\, \frac{\hat{\mathbf{v}} \otimes \hat{\mathbf{v}}}{1-(\hat{\mathbf{\Omega}} 
\cdot \hat{\mathbf{v}} )^2 } \{  \;\sin(2\pi f \tau) -i \,(\hat{\mathbf{\Omega}} \cdot \hat
{\mathbf{v}})( e^{-2 \pi i f \tau \,\hat{\mathbf{\Omega}} \cdot \hat{\mathbf{v}}}-\cos(2\pi f 
\tau)) \} \nonumber \\
&&\;\,-\frac{\hat{\mathbf{u}} \otimes \hat{\mathbf{u}}}{1-(\hat{\mathbf{\Omega}} \cdot \hat
{\mathbf{u}})^2} \{  \;\sin(2\pi f \tau) -i \,(\hat{\mathbf{\Omega}} \cdot \hat{\mathbf{u}})(e^{-2 \pi i f \tau \,\hat{\mathbf{\Omega}} \cdot \hat{\mathbf{u}}}-\cos(2\pi f \tau)) \} \biggr]\;,
\label{eq14}
\end{eqnarray}
where $\tau \equiv L/c$, $\omega$ is the angular frequency of laser, $\vec{\mathbf{X}}$ 
is a position vector of the mirror $M_1$, $R_F$ is the amplitude 
reflectivity of a front mirror, and $R_E$ is the composite amplitude 
reflectivity of the other three mirrors in the cavity. $T_F$ is given by $T_F^2=1-R_F^2$ since we are assuming none of the mirrors have losses. The unit vectors $\hat{\mathbf{u}}$, $\hat
{\mathbf{v}}$ and $\hat{\mathbf{\Omega}}$ are the same as those in Sec. \ref{sec2}. 
Here we define an arm response function
\begin{eqnarray}
{\cal{T}}(f,\,  \hat{\mathbf{\Omega}} \cdot \hat{\mathbf{u}})&\equiv &\frac{- \,e^{-2\pi i f \tau 
}}{2\pi f \tau \, \{ 1-(\hat{\mathbf{\Omega}} \cdot \hat{\mathbf{u}})^2 \}} \biggl[ \,\sin(2\pi f 
\tau )-i\,(\hat{\mathbf{\Omega}} \cdot \hat{\mathbf{u}}) \bigl\{e^{-2\pi i f \tau \,\hat{\mathbf
{\Omega}} \cdot \hat{\mathbf{u}}}-\cos(2\pi f \tau)\bigr\} \biggr] \:, \nonumber \\
&& \label{eq15}
\end{eqnarray}
so that ${\cal{T}}$  gives unity in low frequency limit, as shown in Fig. \ref{fig6}. Using this response function, we 
can rewrite Eq. (\ref{eq14}) in a simple form, 
\begin{eqnarray}
\tilde{\delta \phi}(f, \hat{\mathbf{\Omega}} ) &=& 2\, \omega \tau \,e^{-2\pi i f \hat{\mathbf
{\Omega}} \cdot \vec{\mathbf{X}}/c}\,(1-e^{-4\pi i f \tau})  
\nonumber \\
&&\times \sum_p \mathbf{e}^p \tilde{h}_p : \frac{1}{2}\biggl[\, (\hat{\mathbf{u}} \otimes 
\hat{\mathbf{u}}){\cal{T}}(f,\,  \hat{\mathbf{\Omega}} \cdot \hat{\mathbf{u}})-(\hat{\mathbf
{v}} \otimes \hat{\mathbf{v}}){\cal{T}}(f,\,  \hat{\mathbf{\Omega}} \cdot \hat{\mathbf{v}}) 
\biggr]\;. \nonumber \\
&=& 2\, \omega \tau \,e^{-2\pi i f \hat{\mathbf
{\Omega}} \cdot \vec{\mathbf{X}}/c}\,(1-e^{-4\pi i f \tau}) \, \sum_p \tilde{h}_p(f, \hat{\mathbf{\Omega}} )\, F_p(f, \hat{\mathbf{\Omega}}) 
\label{eq16}
\end{eqnarray}
Therefore, comparing Eq. (\ref{eq16}) with Eq. (\ref{eq17}), we obtain  
\begin{eqnarray}
\tilde{\delta \Phi}^{\prime} (f) &\equiv& \int d\hat{\mathbf{\Omega}}\, \tilde{\delta \Phi}
(f, \hat{\mathbf{\Omega}}) \nonumber \\
&= & \kappa (f)\,\alpha (f)\,\tilde{h} (f)\,.
\label{eq19}
\end{eqnarray}
where $\kappa (f) \equiv 2\, \omega \tau (1-e^{-4 \pi i f \tau })$. To identify $\tilde{h} (f)$ and $\tilde{\delta \Phi}^{\prime} (f)$, we incorporate the extra factor $\kappa (f)\,\alpha (f)$ into the noise spectrum\footnote{Note that our derivation of shot noise is slightly different from our previous paper \cite{bib23}, where we have incorporated the angular response to GW into shot noise. However, here an angular averaging is contained in the GW signals.}, that is,
\begin{eqnarray}
\sqrt{P_{{\rm{shot}}}(f)} &=& |\kappa (f)\,\alpha^{\prime} (f)|^{-1} \, \sqrt{P_{{\rm{qnoise}}}(f)} \nonumber \\
&=&  |\kappa (f)\,\alpha^{\prime} (f)|^{-1} \, \sqrt{\frac{2 \hbar \omega}{\eta I_0}}
\label{eq20}
\end{eqnarray}  
where $\hbar$ is the reduced Planck costant, $\eta$ is the quantum efficiency of photodetector, $I_0$ is laser power, and $P_{\rm{qnoise}}$ is the noise due to vacuum fluctuations. We assumed that shot noise is a dominant noise source around $100\,\rm{MHz}$. In Eq. (\ref{eq20}), $\alpha(f)$ is replaced with
\begin{equation}
\alpha ^{\prime} (f ) \equiv \left| \frac{R_E T_F^2}{(1-R_F R_E)(1-R_F R_E \;e^{-8 \pi i f \tau})} \right|\;, 
\end{equation}
since phase shift due to GW has to be converted into photo current at the photo detector by multiplying a constant factor \cite{bib23}. 

Let us summarize this section. We gave detailed description of SRI and identified the specific forms of GW signal and noise. The GW signal $\tilde{h}^{({\rm{SRI}})}(f)$ completely corresponds to $\tilde{h} (f)$ in Sec. II, which can be calculated using ${\cal{T}}$ in Eq. (\ref{eq15}). The power spectrum of shot noise can be calculated with Eq. (\ref{eq20}), given experimental parameters. In the next section, we will investigate the dependence of overlap reduction function $\gamma$ on the geometrical configuration of two SRIs.

\section{dependence of sensitivity on the relative locations between two detectors}
The SNR is significantly influenced by the relative location of two detectors through the overlap reduction function when the 
wavelength of GW is comparable to the size of a detector. In this section, we will analyze 
the  overlap reduction function in detail and investigate the optimal locations of two 
detectors in an experiment for the detection of GWB at $100\, \rm{MHz}$. 

\begin{figure}[t]
\begin{center}
\includegraphics[width=15cm]{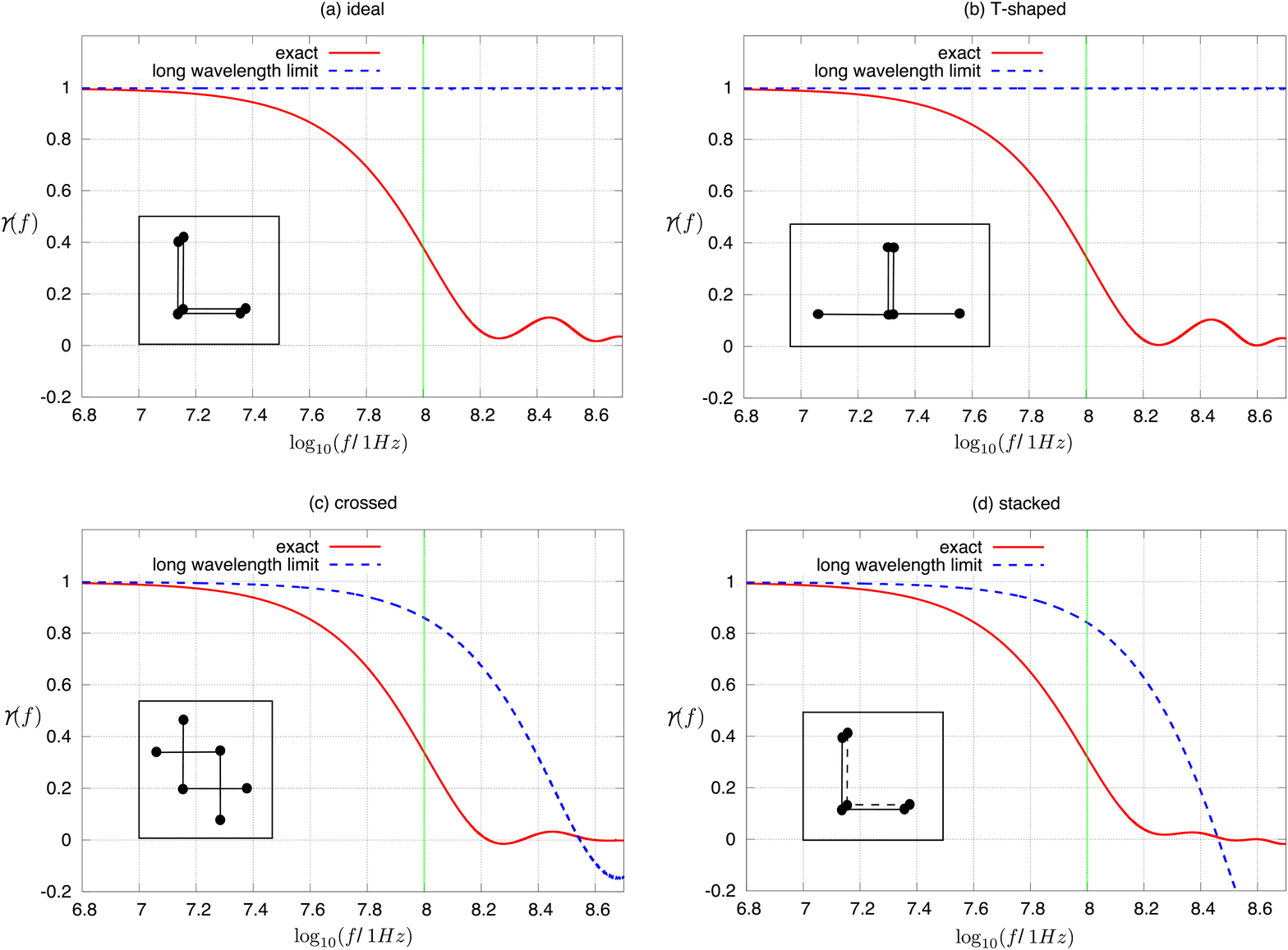}
\caption{(color online) Overlap reduction function in the case of four detector configurations. Each setup is (a) ideal, $\Delta \vec{\mathbf{X}}=0$, $\beta=0$, (b) T-shaped, $\Delta \vec{\mathbf{X}}=0$, $\beta=\pi/2$, (c) crossed, $\Delta \vec{\mathbf{X}}=(L/2, L/2, 0)$, $\beta=\pi$, (d) stacked, $\Delta \vec{\mathbf{X}}=(0, 0, L/2)$, $\beta=0$. The "exact" means the calculation with arm response function ${\cal{T}}$ and the "long wavelength limit" 
${\cal{T}}=1$. The latter is not valid around $100\,\rm{MHz}$, but merely plotted for comparison. Note that the sign of $\gamma(f)$ in (b) is inversed for convenience of comparison.}
\label{fig2}
\end{center}
\end{figure}

The overlap reduction function can be calculated numerically from Eq. (\ref{eq8}) using 
the arm response function ${\cal{T}}$ given in Eq. (\ref{eq15}), where the phase factor $e^{2\pi i f 
\hat{\mathbf{\Omega}} \cdot \Delta \vec{\mathbf{X}}/c }$ plays an important role. To see 
this, we consider the four configurations of detectors and calculate $\gamma (f)$. The results are shown in Fig. \ref{fig2} with the case of "exact" and "long wavelength limit". The former is calculated with the full arm response function ${\cal{T}}$. The latter is calculated with ${\cal{T}}=1$, which is just plotted for reference, though the approximation is not valid around $100\,\rm{MHz}$. Each configuration of detectors is characterized by the relative position $\Delta \vec{\mathbf{X}} = \vec{\mathbf{X}}
_1-\vec{\mathbf{X}}_2$ and the relative angle $\beta$. Note that $\vec{\mathbf{X}}_i$ is the position vector of $M_1$ of i-th detector (see Fig. \ref{fig1}). In the case of (a) ideal, $\gamma(f)$ rapidly decreases even though the detectors are completely colocated and coaligned, because we defined the $\gamma (f)$ including arm response functions. Case (b) T-shaped has a behavior similar to that of (a) for the same reason. The arm response function is needed to take the effect of the phase change of GW at high frequencies into account. Cases (a) and (b) are similar, but have a subtle difference since the arms of detectors are at different locations and experience different phases of GW. As a result, the overlap of case (b) is a little worse than case (a). In the cases of (c) crossed and (d) stacked, $\gamma (f)$ also decrease more rapidly than in the long wavelength limit. This is because the contribution of the GW phase change at high frequencies is added to that in the long wavelength limit. Therefore, we cannot obtain $\gamma (f)=1$ at $100\,\rm{MHz}$ with the detectors where detector size and GW wavelength are comparable. It follows that SNR is worsened by a factor of $(0.377)^{-1} \approx 2.65$ at $100\,\rm{MHz}$ in contrast to the case where long wave approximation is valid, even if the two detectors' configuration is optimal. One may expect unit response to be obtained by constructing much smaller detectors. However, the total response of detectors $\tilde{\delta \Phi}$ is worsened since the resonant frequency of GW signal also depends on the detector size and is shifted upward. Thus, this loss of sensitivity due to the phase change of GW is inevitable.

\begin{figure}[h]
\begin{center}
\includegraphics[width=8cm]{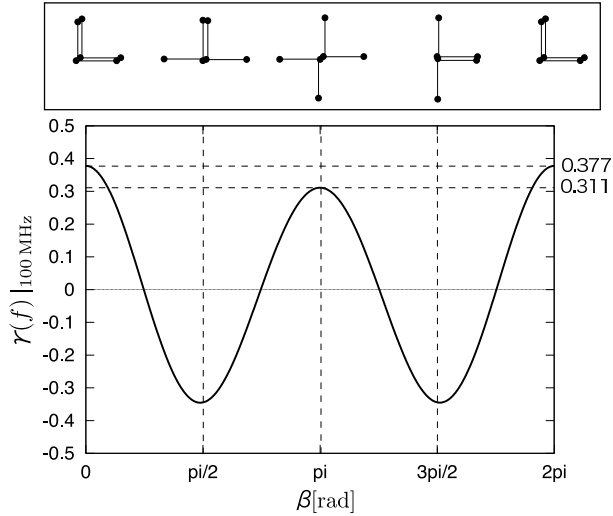}
\caption{Overlap reduction function when the detector is initially colocated and coaligned and is rotated at the same location. $\beta$ is the rotation angle. }
\label{fig3}
\end{center}
\end{figure}

\begin{figure}[h]
\begin{center}
\includegraphics[width=8cm]{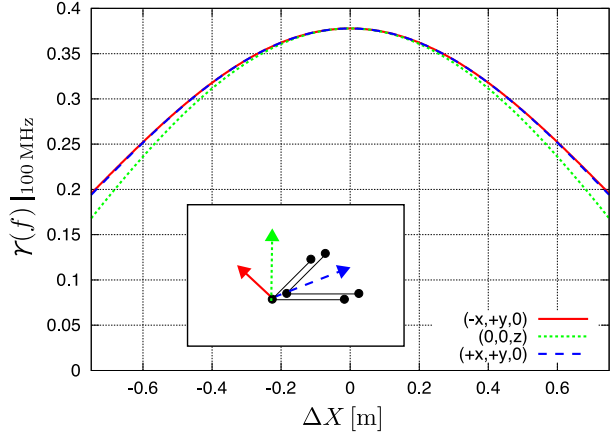}
\caption{(color online) Overlap reduction function when the detector is initially colocated and coaligned ($\beta=0$) and is translated in certain directions. Each curve means the direction of translation. $(+x, +y, 0)$ is the direction of $(\hat{\mathbf{u}}+\hat{\mathbf{v}})/\sqrt{2}$, $(-x, +y, 0)$ is the direction of $(\hat{\mathbf{u}}-\hat{\mathbf{v}})/\sqrt{2}$, and $(0, 0, z)$ is the direction perpendicular to the $\hat{\mathbf{u}}\,\hat{\mathbf{v}}$ plane. }
\label{fig4}
\end{center}
\end{figure}

\begin{figure}[h]
\begin{center}
\includegraphics[width=8cm]{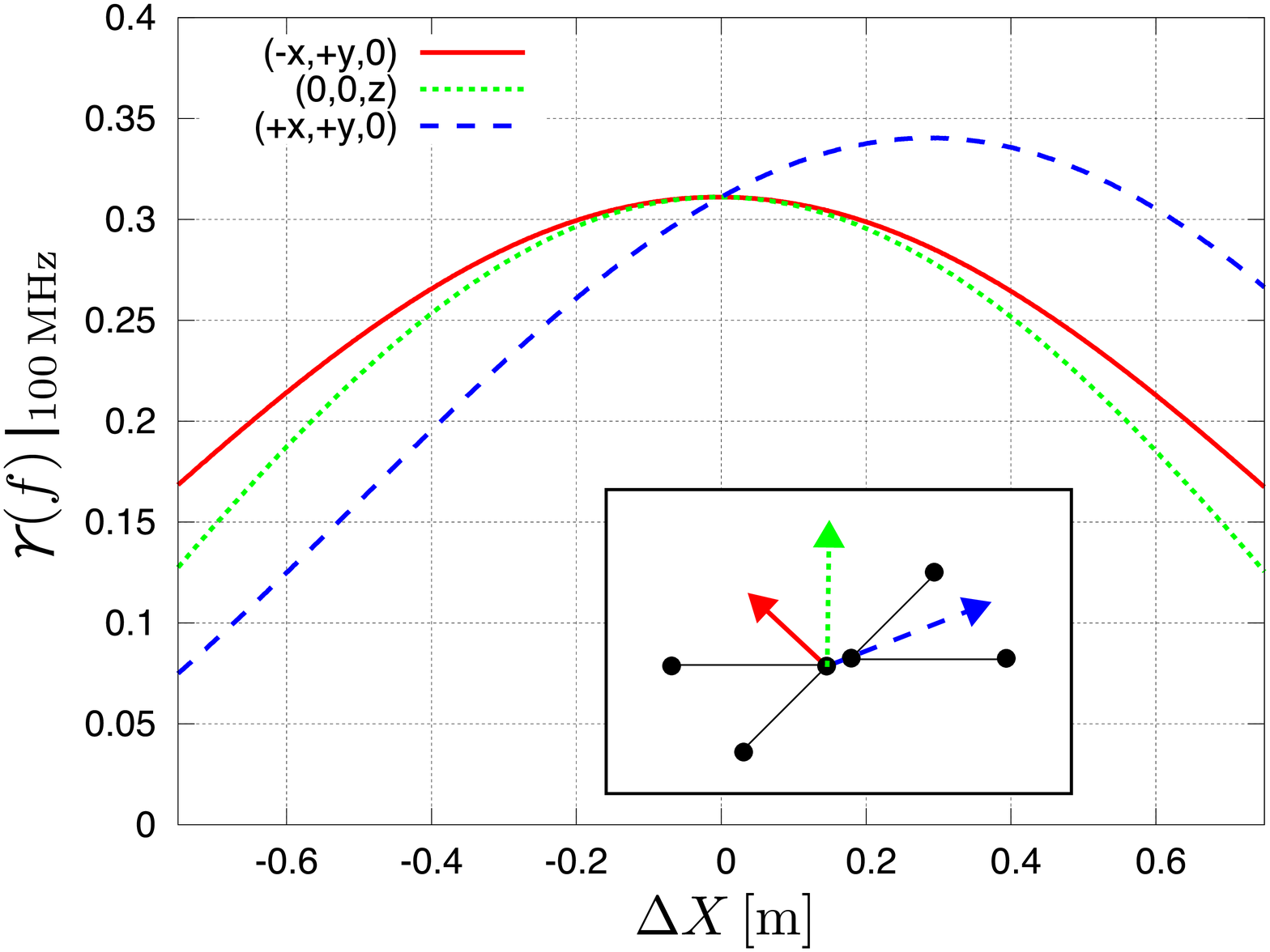}
\caption{(color online) Overlap reduction function when the detector is initially colocated and reversed ($\beta=\pi$) and is translated in certain directions. Each curve means translation in the same direction as shown in Fig. \ref{fig4}.}
\label{fig5}
\end{center}
\end{figure}

Next, we will fix the frequency at $100\,\rm{MHz}$ and consider $\gamma(f)$. In fact, SRI has a narrow frequency band and what we are most interested in is $\gamma$ at $100\,\rm{MHz}$. In Fig. \ref{fig3}, the location of one detector is fixed, while the other detector is located at the same site ($\Delta \vec{X} =0$) and the directions of arms are rotated. In this case, the magnitude of $\gamma(f)$ oscillates, however, it has the maximum peak at $\beta =0$ and the minimum peak at $\beta =\pi$. It is intuitive that the GW signal is better correlated when the directions of arms are coaligned. In Figs. \ref{fig4} and \ref{fig5}, the angle of detectors is fixed and the locations are translated. In an initially coaligned case ($\beta=0$) in Fig. \ref{fig4}, as expected, $\gamma(f)$ has its maximum at $\Delta X =0$ and keeps the moderate value in the range of $\Delta X =\pm 0.2 \,\rm{m}$. In an initially reversed case ($\beta=\pi$) in Fig. \ref{fig5}, an interesting feature can be seen. When the detector is translated to the direction $(\hat{\mathbf{u}}+\hat{\mathbf{v}})/\sqrt{2}$, the peak of $\gamma(f)$ is shifted. This is because the overlap of the two detectors is better when their arms are overlapped geometrically like (c) in Fig. \ref{fig2}.

\section{Sensitivity to GWB}
We will describe what the best location of detectors with respect to the sensitivity is, and calculate the sensitivity achievable with correlation analysis. From the results obtained above, the best location is obviously colocated and coaligned, and gives $\gamma(f)|_{100\,\rm{MHz}} \approx 0.377$. As shown in Fig. \ref{fig4}, this value is hardly changed in the range of $\Delta X =\pm 0.2 \,\rm{m}$ for coaligned detectors. In an experiment, it is impossible to put the detectors in completely colocated and coaligned location because of the restricted experimental space of the optics. However, experimental detector configuration does not significantly affect $\gamma(f)$ if the detectors are nearly colocated and coaligned. Therefore, we fix it to $\gamma_{\rm{opt}} = 0.377$. 

As for the power spectral density of noise in an SRI, one can calculate from Eq. (\ref{eq20}). We select the arm length of a detector as $L=0.75\,\rm{m}$ so that the GW signal resonates at $100\,\rm{MHz}$. With experimental parameters, $\omega =1.77\times 10^{15}\, \rm{rad\: s}^{-1}$ and $\eta=1$, the power spectral density of noise \footnote{Note that this is not the squared strain-amplitude noise ordinarily used in the noise curve of ground-based detectors. To convert, one has to multiply Eq. (\ref{eq21}) by $(2 \pi f \tau / \sin 2\pi f \tau)^2$, in this paper, which is incorporated into the GW signal.} around $100\,\rm{MHz}$ is
\begin{equation}
P_i (f) \approx 4.65 \times 10^{-42} \biggl( \frac{1.60 \times 10^4}{\alpha ^{\prime}(f)} \biggr)^2 \biggl( \frac{1\rm{W}}{I_0} \biggr) \,\rm{Hz}^{-1}, \;\;\;\;i=1,2.
\label{eq21}
\end{equation}
The factor $\alpha^{\prime}$ is called the optical amplification factor in a cavity and gives $\alpha ^{\prime} \approx 1.6 \times 10^4$ with the reflectivity of the recycling mirror, $R_F^2=0.99996$, and the reflectivity of the other three mirrors, $R_E^2=(0.99998)^3$. The bandwidth is $\sim 2 \,\rm{kHz}$ with these reflectivities. Substituting $P_i (f)$ and $\gamma_{\rm{opt}}$ into Eq. (\ref{eq18}), and assuming that observation time is $T=1\,\rm{yr}$ and that $\Omega_{\rm{gw}}(f)$ has a flat spectrum around $100\,\rm{MHz}$ (which is sufficient for practical purposes \cite{bib28}), one can calculate the sensitivity of two SRIs to GWB and obtain $h_{100}^2 \Omega_{\rm{gw}} \approx 1.4 \times 10^{14}$. 
 
\section{conclusions}
In this paper, we investigated the optimal location of two SRIs whose arms are $0.75\,\rm{m}$, and derived its sensitivity
to GWB in correlation analysis. At $100\,\rm{MHz}$, the wavelength of GW is comparable to the size of a detector. This means that the GW response of the detector is less effective than one in the long wavelength limit, and that the location of detectors significantly affects the sensitivity. We included the effect due to the finite size of a detector into the overlap reduction function and evaluated it. As a result, SNR is worsened by a factor of $(0.377)^{-1} \approx 2.65$ at $100\,\rm{MHz}$ in contrast to the case where long-wave approximation is valid, even if the two detectors' configuration is optimal. This is because
the phase of GW changes during light's roundtrip in an arm. SNR also depends on the relative distance and angle between detectors. However, SNR is almost optimal value if two detectors are in the range of $\pm 0.2\, \rm{m}$ and coaligned. Using this configuration and experimentally achieved parameters of two SRIs, one can achieve the sensitivity to GWB, $h_{100}^2 \Omega_{\rm{gw}} \approx 1.4 \times 10^{14}$ corresponding to the amplitude $h \sim 10^{-23}\,\rm{Hz}^{-1/2}$. This constraint on GWB would be much tighter than that obtained by current direct observation, $h \sim 10^{-14}\,\rm{Hz}^{-1/2}$ \cite{bib22}. However, to reach indirect constraint on GWB by big-bang nucleosynthesis, further improvement of the sensitivity is required.

\begin{acknowledgments}
This research was supported by the Ministry of Education, Science, Sports and Culture, 
Grant-in-Aid for Scientific Research (A), 17204018.
\end{acknowledgments}

\bibliography{UHFGW2}

\end{document}